\documentclass{JINST}

\usepackage{graphicx}

\title{Simulation of beam-beam effects and Tevatron experience}

\author{Alexander Valishev$^a$\thanks{Corresponding
author.}, Yuri Alexahin$^a$, Valeri Lebedev$^a$~ and Dmitry Shatilov$^b$\\
\llap{$^a$}Fermi National Accelerator Laboratory,\\
  PO Box 500, Batavia, IL, 60510 USA\\
\llap{$^b$}Budker Institute of Nuclear Physics,\\
  Novosibirsk, 630090, Russia\\
  E-mail: \email{valishev@fnal.gov}}

\abstract{
Effects of electromagnetic interactions of colliding bunches in the Tevatron
had a variety of manifestations in beam dynamics presenting
vast opportunities for development of simulation models and tools. 
In this paper the computer code for simulation of weak-strong beam-beam effects
in hadron colliders is described. We report the collider operational experience
relevant to beam-beam interactions,
explain major effects limiting the collider performance and compare results of
observations and measurements with simulations.}


\begin{document}

\section{Introduction}

Peak luminosity of the Tevatron reached 4.3$\times 10^{32}$ cm$^{-2}$s$^{-1}$,
which exceeds the original Collider Run II goal \cite{TevJINST}. 
This achievement became possible due to numerous upgrades in the antiproton
source, injector chain, and in the Tevatron collider itself. 
One of the most notable rises of luminosity came from the commissioning of
electron cooling in the Recycler Ring and advances in the antiproton
accumulation rate \cite{eCool}.
Starting from 2007, the intensity and brightness of antiprotons delivered to
the collider greatly enhanced the importance of beam-beam effects. Several
configurational and operational improvements in the Tevatron have been planned
and implemented in order to alleviate these effects and allow stable running at
high peak luminosities.

Since the publication of paper \cite{BBTev} that gave a detailed summary of beam
dynamics issues related to beam-beam effects, the peak luminosity of Tevatron
experienced almost a tree-fold increase. In the present article we provide an
updated view based on the last years of collider operation
(Section \ref{sec-overview}).

Development of a comprehensive computer simulation of beam-beam effects in the
Tevatron started in 1999. This simulation proved to be a useful tool for
understanding existing limitations and finding ways to mitigate them.
As the first step, we developed a simplified model of beam parameter evolution
in the Tevatron, which includes all important phenomena affecting luminosity
decay, excluding the beam beam effects (Section \ref{sec-storeanalysis}).
For the numerical simulations of beam-beam effects, we used a multiparticle
tracking code Lifetrac.
In Section \ref{sec-lifetrac} the main features of the code are
described. In Sections \ref{sec-colhelix}-\ref{sec-C2} we summarize our
experience with simulations of beam-beam effects in the Tevatron, and
cross-check the simulation results against various experimental data and
analytical models. We also correlate the most notable changes in the machine
performance to changes of configuration and beam conditions, and support the
explanations with simulations.

\section{\label{sec-overview}Overview of beam-beam effects}

A detailed description of the Tevatron collider Run II can be found elsewhere
\cite{TevJINST}. Here we provide only essential features important for
understanding of the beam dynamics.

The Tevatron was a superconducting proton-antiproton collider ring in which
beams of the two species collided at the center of mass energy of
$2 \times 0.98$ TeV at two experiments. Each beam consisted of 36 bunches
grouped in 3 trains of 12 with 396 ns bunch spacing and 2.6 $\mu$s abort gaps
between the trains. The beams shared a common vacuum chamber with both beams
moving along helical trajectories formed by electrostatic separators.
Before the high energy physics collisions
have been initiated, the proton and antiproton beams could be moved
longitudinally with respect to each other, which is referred to as cogging.
This configuration allowed for 72 interactions per bunch each turn with the
total number of collision points in the ring equal to 138.
The total number of collision points was determined by the symmetry of bunch
filling pattern.

At the peak performance Tevatron operated with approximately
$N_p = 2.8 \cdot 10^{11}$
protons and $N_a = 0.9 \cdot 10^{11}$ antiprotons per bunch at the beginning of
a store. The normalized transverse 95\% beam emittances were
$\varepsilon _p = 18 \cdot 10^{-6}$m for protons and
$\varepsilon _a = 7 \cdot 10^{-6}$m for antiprotons. 
The rms length of proton and antiproton bunches at the beginning of a
high energy physics
(HEP) store was 52 cm and 48 cm, respectively. Parameters of the beams were
mostly determined by the upstream machines.

The value of $\beta$-function at the main collision points ($\beta ^*$) was
0.28 m. Betatron tunes were $Q_x = 20.584$, $Q_y = 20.587$ for protons and
$Q_x = 20.575$, $Q_y = 20.569$ for antiprotons.

\begin{figure}[htb]
   \includegraphics*[width=\textwidth]{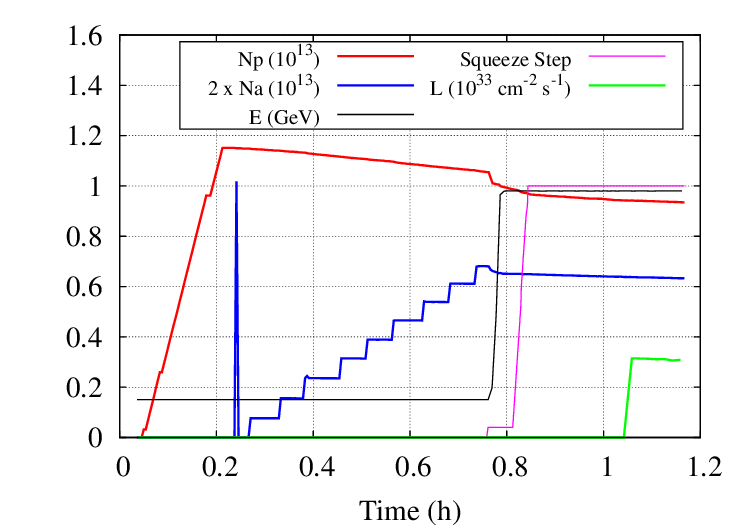}
   \caption{\label{store5989}Collider fill cycle for store 5989.}
\end{figure}

A typical collider fill cycle is shown in Figure \ref{store5989}. First, proton
bunches were injected one at a time on the central orbit. After that, the
electrostatic separators were powered and antiproton bunches were injected in
batches of four. This process was accompanied by longitudinal cogging after each
3 transfers. Then the beams were accelerated to the top energy in 85 s and the
machine optics was changed to collision configuration in 25 steps over 120
seconds of the so-called low-beta squeeze, during wich the $\beta ^*$ changed
from 1.5 to 0.28 m.
The last two stages included initiating collisions
at the two main interaction points (IP) and removing halo by moving in
the collimators.

\begin{figure}[htb]
   \includegraphics*[width=\textwidth]{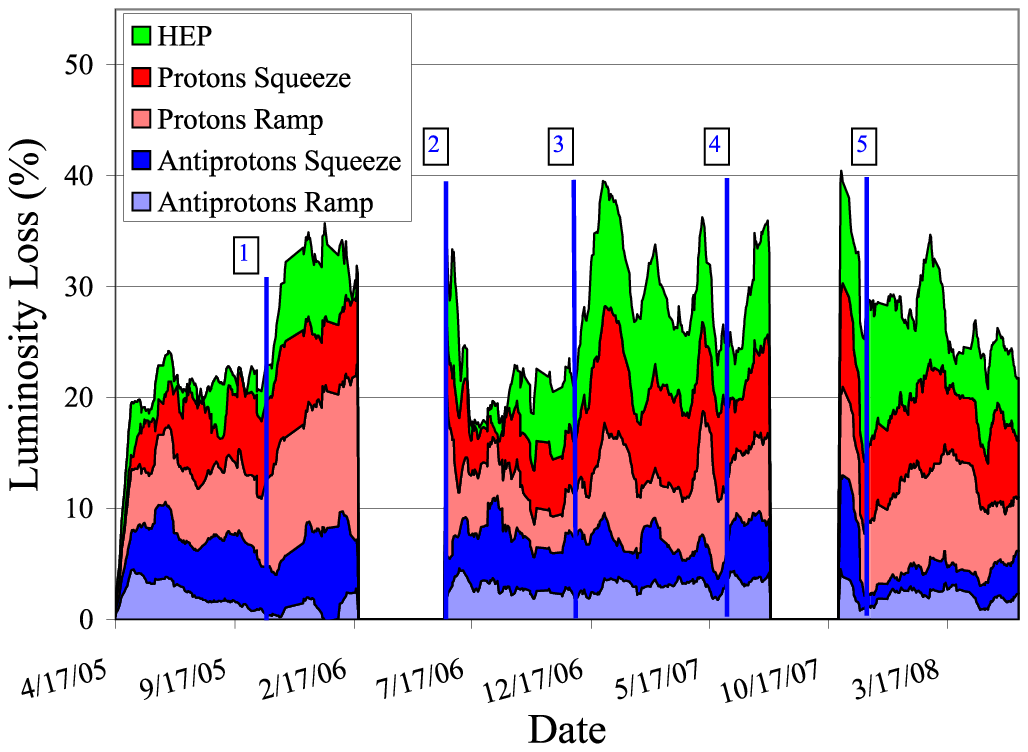}
   \caption{\label{lossbudget}Luminosity loss budget over a 3 year period.
   The labels mark: 1. Commissioning of electron cooling. 2. Installation of
   extra separators and new collision helical orbit. 3. Improved antiproton
   accumulation rate (stochastic cooling upgrade).
   4. Correction of second-order chromaticity. 5. Implementation of controlled
   antiproton emittance blowup.}
\end{figure}

It has been shown in machine studies that beam losses up the ramp and through
the low-beta squeeze were mainly caused by beam-beam effects \cite{BBTev}.
During the high energy physics collisions runs (HEP mode),
the beam-beam induced emittance growth and particle losses
contributed to the faster luminosity decay. Figure \ref{lossbudget} summarizes
the observed losses of luminosity during different stages of the collider cycle.

\subsection{Beam-beam effects at injection energy}
\begin{figure}[htb]
   \includegraphics*[width=\textwidth]{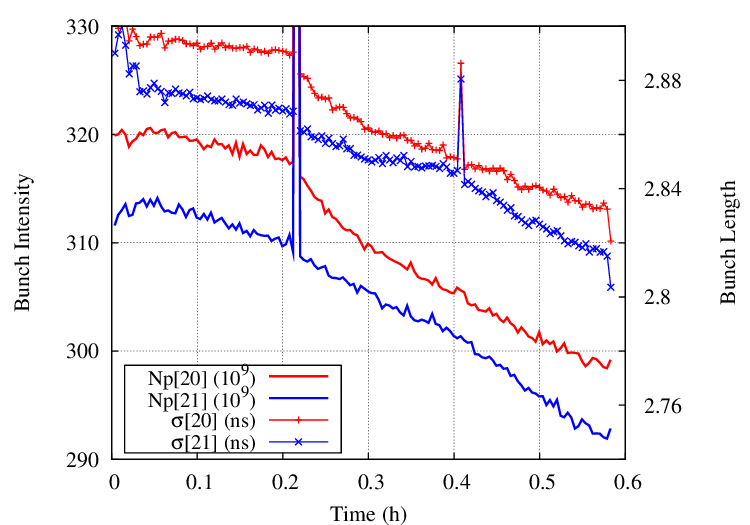}
   \caption{\label{bunch2021}Intensity and rms length (ns)
   of proton bunches no. 20 and
   21 during injection of antiprotons.}
\end{figure}

During injection the long range (also referred to as parasitic) beam-beam
effects caused proton losses (usually 5 to 10\%). At the same time the
antiproton lifetime $(dN_a/dt/N_a)^{-1}$ was very good and only a fraction of
a per cent were lost.
Observations showed that mainly off momentum particles were lost
(Figure \ref{bunch2021}) and the betatron tune chromaticity $C = dQ/d \delta$,
where $\delta = \Delta p/p$ is the relative momentum deviation, had a remarkable
effect. Early in Run II the chromaticity had to be kept higher than 8 units in
order to maintain coherent stability of the intense proton beam, but after
several improvements aimed at reduction of the machine impedance the
chromaticity was about 3 units \cite{headtail03,headtail05,headtail08}. 
Figure \ref{bunch2021} shows an interesting feature in the behavior of two
adjacent proton bunches (no. 20 and 21). Spikes in the measured values are
instrumental effects labeling the time when the beams are cogged. Before the
first cogging the bunches have approximately equal lifetime. After the first
cogging bunch 20 exhibits faster decay, and bunch 21 after the second. Analysis
of the collision patterns for these bunches allowed to pinpoint a particular
collision point responsible for the lifetime degradation. The new optimized
helical orbit separation scheme at injection energy
has been implemented late in 2007 which improved the proton lifetime 
\cite{helix05,YAHelices}.

\subsection{Beam-beam effects during low-beta squeeze}
During the low-beta squeeze two significant changes occured - the $\beta^*$
value was being gradually decreased from $\sim$1.5 m to 0.28 m (hence the
name squeeze) and the helical orbits changed their shape and polarity from
injection to collision configuration. The latter posed a serious limitation
since the beams separation at several long range collision points briefly
decreased from 5-6$\sigma$ to $\sim$2$\sigma$. At this moment a sharp spike
in losses was observed. 

Another important operational concern was the tight aperture limitation in one
of the two final focus regions (CDF). With dynamically changing orbit and
lattice parameters the local losses were often high enough to cause a quench of
the superconducting magnets even though the total amount of beam loss was small
($\sim$ 1\%). The aperture restriction has been located and fixed in October
of 2008.

\begin{figure}[htb]
   \includegraphics*[width=\textwidth]{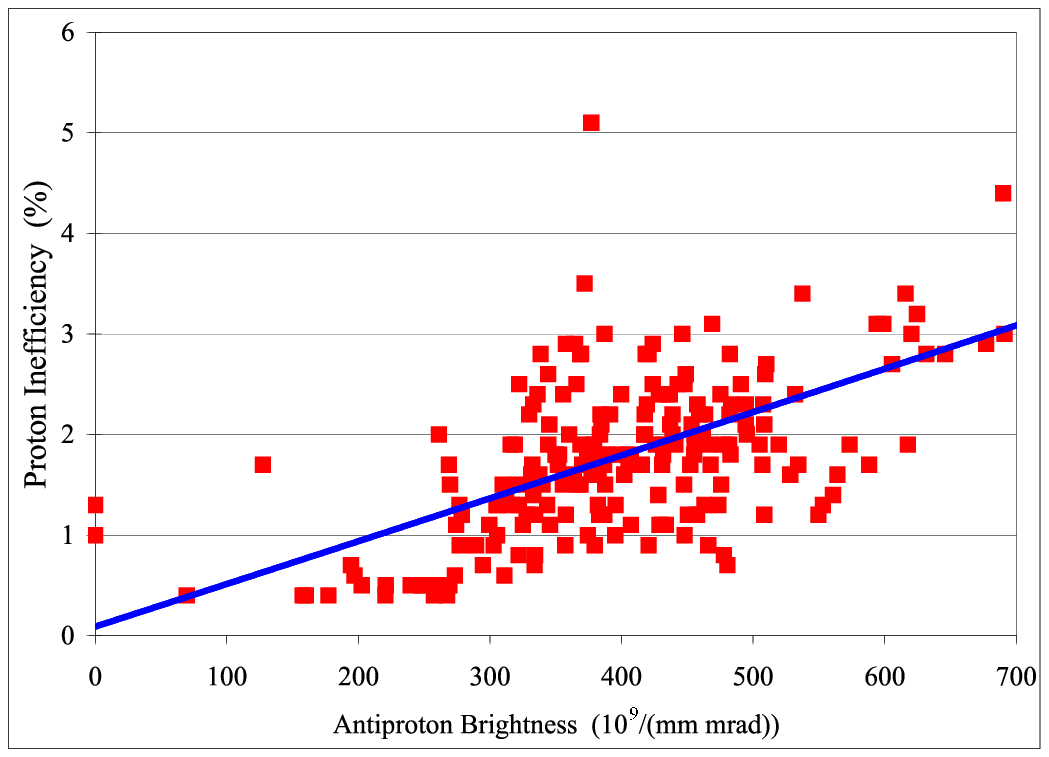}
   \caption{\label{ploss}Proton losses during low-beta squeeze vs. 
antiproton beam brighness $36 \cdot N_a /\varepsilon _a$.}
\end{figure}

Besides orbit stability two other factors were found to be important in
maintaining low losses through the squeeze: antiproton beam brightness and
betatron coupling. Figure \ref{ploss} shows the dependence of proton losses on
the antiproton beam brightness. Improvements implemented in the antiproton
source during the 2007 shutdown resulted in the increase of antiproton beam
brightess. This, in turn, caused a large number of stores to be lost at that
stage of the Tevatron cycle, and hence demanded the commissioning of antiproton
emittance control system \cite{PBJpaper}.

\subsection{Beam-beam effects during high energy physics operation}
After the beams were brought into collisions at the main IPs, there were two
head-on and 70 long range collision points per bunch. Beam-beam effects caused
by these interactions lead to emittance growth and particle losses in both beams. 

During the running prior to the 2006 shutdown the beam-beam effects at HEP
mostly affected antiprotons. The long range collision points nearest to the main
IPs were determined to be the leading cause for poor lifetime. Additional
electrostatic separators were installed in order to increase the separation at
these IPs from 5.4 to 6$\sigma$ \cite{YAHelices}. Also, the betatron tune
chromaticity was decreased from 20 to 10 units. Since then, the antiproton
lifetime was dominated by losses due to luminosity and no emittance growth was
observed provided that the betatron tune working point was well controlled.

\begin{figure}[htb]
   \includegraphics*[angle=-90,width=\textwidth]{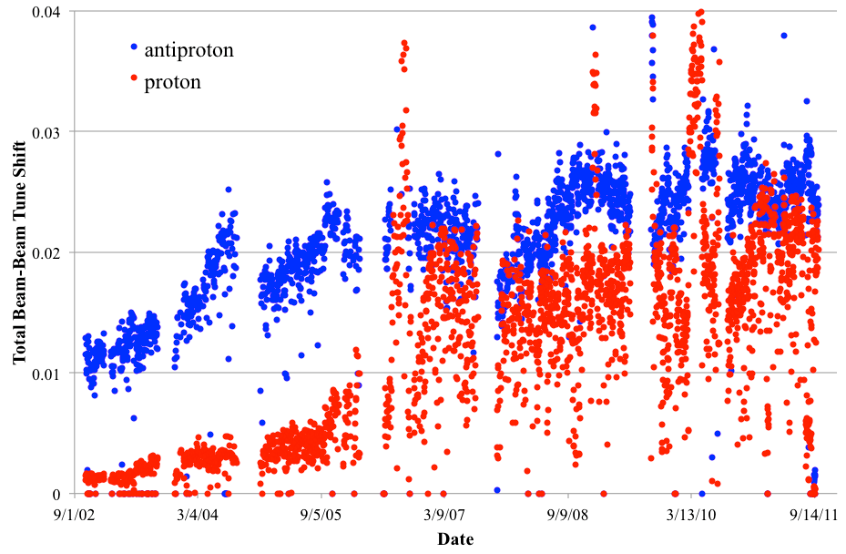}
   \caption{\label{xi_store}Head-on beam-beam tune shift parameter vs. time.}
\end{figure}

Commissioning of electron cooling of antiprotons in the Recycler resulted in
smaller emittances, which together with increased antiproton
stacking rate drastically changed the situation for protons. Figure
\ref{xi_store} shows the evolution of total head-on beam-beam tune shift
$\xi = N r_p / 4 \pi \epsilon$ (where $N$ is the number of particles,
$r_p$ is the classical proton radius, $\epsilon$ is the normalized rms emittance)
for protons and antiprotons. Note that prior to the 2006 shutdown the proton
$\xi$ was well under 0.01 and big boost occurred in 2007 when both beam-beam
parameters became essentially equal. It was then when beam-beam related losses
and emittance blowup started to be observed in protons.

\begin{figure}[htb]
   \includegraphics*[width=\textwidth]{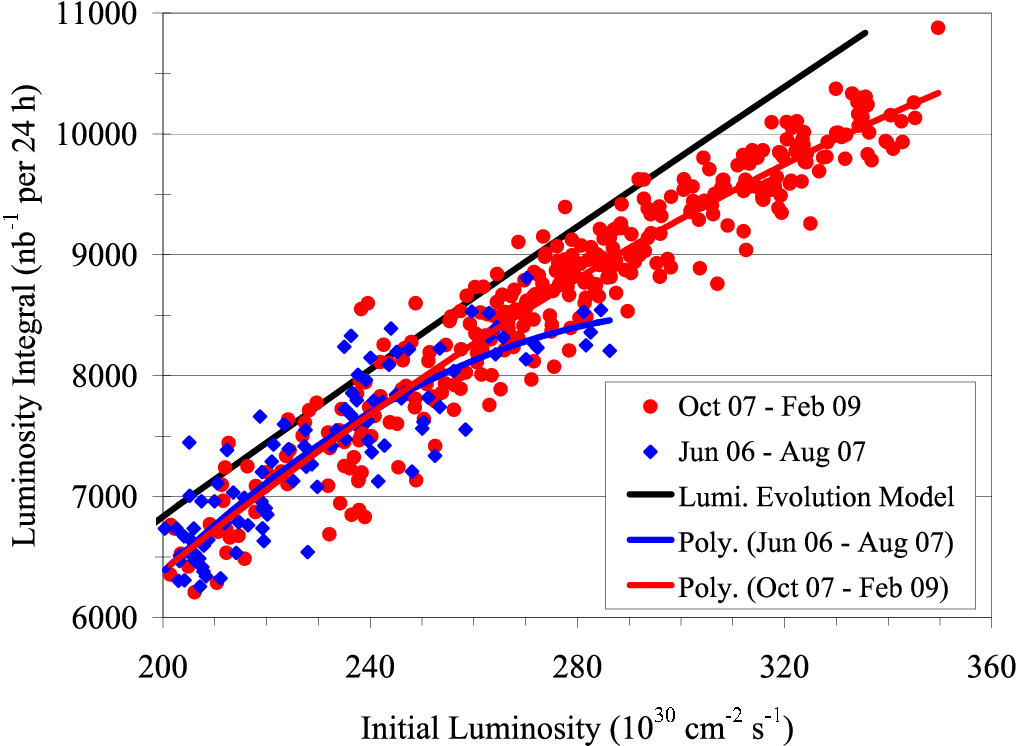}
   \caption{\label{lumi_int}Luminosity integral normalized for 24 hrs store vs.
   initial luminosity. Blue points and curve - before the second order chromaticity
   correction, red - after the correction. Black line represents the maximum
   possible luminosity integral for the given beam parameters in the absence of
   beam-beam effects.}
\end{figure}

Our analysis showed that deterioration of the proton lifetime was caused by a
decrease of the dynamical aperture for off-momentum particles due to head-on
collisions (see Sec. \ref{sec-C2}). It was discovered that the Tevatron optics
had large chromatic perturbations, e.g. the value of $\beta^*$ for off-momentum
particles could differ from that of the reference particle by as much as 20\%.
Also, the high value of second order betatron tune chromaticity
$d^2 Q/ d (\Delta p/p)^2$ generated a tune spread of $\sim$0.002.
A rearrangement of
sextupoles in order to correct the second order chromaticity has been planned and
implemented in 2007 shutdown \cite{C2Comp}. Figure \ref{lumi_int}
demonstrates the effect of this modification on the integrated luminosity.
The time dependence of luminosity is very well approximated by $L_0/(1+t/\tau)$.
Therefore one can normalize the luminosity integral for a given store to a fixed
length $T_0$ by using the expression
$L_0 \tau \cdot ln(1+T_0/\tau)$ \cite{lumevol}.
Here $L_0$ is the initial luminosity, and $\tau$ is the luminosity lifetime.
One can see that after the modification the saturation at luminosities above
$2.6 \times 10^{32}$cm$^{-2}$s$^{-1}$ was mitigated and the average luminosity
delivered to experiments increased by some 10\%.

Another increase in the proton $\xi$ occurred after the 2007 shutdown when the
transverse antiproton emittance was decreased due to improvements
in the injection optics matching.
The total attained head-on beam-beam tune shift for protons became higher than
that of antiprotons and exceeded 0.03. That led to high sensitivity of the
proton lifetime to minor variations of the betatron tunes, and to severe
background conditions for the experiments. The reason was believed to be the
large betatron tune spread generated by significantly unequal beam sizes at the
IPs \cite{syphers}. Indeed, at times the antiproton beam size was a
factor of 2 to 2.5 smaller than the proton size.

A dedicated system has been commissioned to reduce the proton-to-antiproton
emittance ratio. It increased the antiproton emittance after the top energy was
reached by applying wide band transverse noise to a directional strip line
(line 5 in Fig. \ref{lossbudget}) \cite{PBJpaper}.
Ultimately, the optimal emittance ratio for operations was determined to be
about 3 (i.e. the factor of 1.7 in beam sizes).

The remaining part of this paper deals with beam-beam effects in the HEP stores.
Discussions on the long range effects at the injection energy and coherent
beam-beam effects \cite{YAbb} are left out of
the scope of this report.

\section{\label{sec-storeanalysis}Multi-process beam physics analysis}
\begin{figure}[htb]
   \includegraphics*[width=\textwidth]{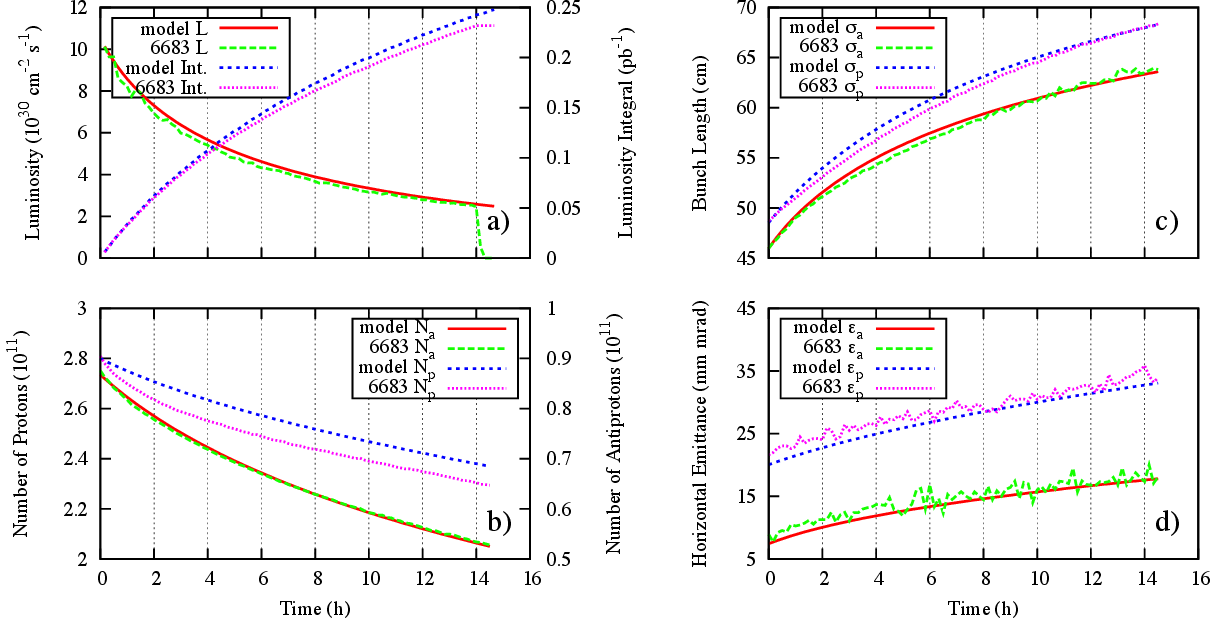}
   \caption{\label{storea6683}Observed beam parameters in store 6683 compared
to store analysis calculation (model). $L_0=3.5\cdot 10^{32}$cm$^{-2}$s$^{-1}$. 
a) single bunch luminosity and luminosity integral. b) intensity of proton
bunch no. 6 and of antiproton bunch colliding with it (no. 13).
c) rms bunch lengths. d) horizontal 95\% normalized bunch emittances.}
\end{figure}

The beam-beam interaction was not the single strongest effect determining the
evolution
of beam parameters at collisions. There were other sources of diffusion causing
the emittance growth and particle losses, including the intrabeam
scattering, noise of accelerating RF voltage, ground motion,
scattering on the residual gas, and particle burnup due to luminosity
(non-elastic scattering).
Parameters of these mechanisms were measured in beam studies (see e.g.
\cite{IBS_JINST}), and then a model
was built in which the equations of diffusion and other processes were solved
numerically \cite{val03}. This model was able to predict evolution of the beam
parameters in the case of weak beam-beam effects. When these effects are not
small, it provides a reference for evaluation of their strength. 
We used that analysis on a store-by-store basis to monitor the machine 
performance in real time \cite{storeanalysis} because such calculations were
very fast compared to a full numerical beam-beam simulation.
Figure \ref{storea6683} presents an example comparison of the evolution of beam
parameters in an actual high luminosity store to the calculations. Note that there
is no transverse emittance blow up in both beams, and the emittance growth is
dominated by processes other than beam-beam interaction. The same is true for
antiproton intensity and bunch length. The most pronounced difference between
the observation and the model is seen in the proton intensity evolution.
Beam-beam effects
caused the proton lifetime degradation during the initial 2-3 hours of the store
until the proton beam-beam tune shift drops from 0.02 to 0.015.
The corresponding loss of the luminosity integral was about 5\%.

\section{\label{sec-lifetrac}Numerical simulations of beam-beam effects}

Numerical simulations of beam-beam effects in the Tevatron were carried out
with the use of weak-strong particle tracking code Lifetrac.
Initially this code was developed for simulation of the
equilibrium distributions of the particles in circular electron-positron
colliders \cite{lifetrac96}. In 1999 the new features have been implemented,
which allowed simulating non-equilibrium distributions, for example in proton
beams. In this case the goal of simulations is not to obtain the equilibrium
distribution but to observe the evolution of initial distribution.
The number of simulated particles can vary in the range from $10^3$
to $10^6$, usually it is set to $(5 \div 10) \cdot 10^3$. The tracking time
is divided into ``steps'', typically $10^3 \div 10^5$ turns each.
The statistics obtained
during the tracking (1D histograms, 2D density in the space of normalized
betatron amplitudes, luminosity, beam sizes and emittances) is averaged over all 
particles and all turns for each step. Thus, a sequence of frames
representing evolution of the initial distribution is obtained.

Another important quantity characterizing the beam dynamics is the intensity
lifetime. It is calculated by placing an aperture restriction in the machine
and counting particles reaching the limit. The initial and final coordinates of
the lost particle are saved. This information is valuable for analysis of
various beam dynamics features.

The initial 6D distribution of macroparticles can be either Gaussian
(by default), or read from a separate text file. The macroparticles may also
have different ``weights''. This allows representing the beam tails more
reliably with a limited number of particles. Usually we simulate the Gaussian
distribution with weights: particles initially located in the core region have
larger weights while the ``tail'' particles with smaller weights are more numerous.

In the Tevatron bunch pattern (3 trains of 12 bunches) there are two main
IPs and 70 long range collision points for each bunch. When performing
transformation through a main IP, the ``strong'' bunch is divided into slices
longitudinally. The higher are the orders of significant betatron resonances
which are supposed to make effect on the distribution, the greater must be the 
number of slices. In our simulations 12 slices were used in the main IPs where
beta-functions are approximately equal to the bunch length and only one slice
in long range collision points where beta-functions are much greater and one can 
neglect the betatron phase advance over the bunch length.

The transverse density distributions within ``strong'' slices are bi-Gaussian,
allowing to apply the well-known formulae \cite{sympl} for the 6D symplectic 
beam-beam kick. However, a simple modification allowed simulating non-Gaussian
strong bunches. Namely, the strong bunch is represented as a superposition of
a few (up to three) Gaussian distributions with different betatron emittances.
The kicks from all these ``harmonics'' are summarized additively.
The calculation time is increased somehow (not very significantly) but the
transformation remains 6D symplectic.

\subsection{Tevatron optics}
The parasitic collisions in Tevatron played significant role in the beam
dynamics. In order to account for their contribution correctly, an accurate
knowledge of the focusing optics of the whole ring with all distortions,
beta beatings, coupling, etc. was required. This necessitated the construction
of a realistic model of the machine lattice based on beam measurements.
The most effective method proved to be the orbit responce matrix analysis
\cite{diff_orb,TevOptics05,loco2006}. 

The model lattice was built in the optics code OptiM \cite{optim}.
Both OptiM and Lifetrac treat betatron coupling using the same coupled
beta-functions formalism \cite{lebedev-bogacz}. That allows the linear transport
matrix between any two points to be easily derived from the coupled lattice
functions and phase advances.

A set of scripts has been created enabling fast creation of input files for 
the beam-beam simulation. These programs automate calculation of azimuthal
positions of interaction points for the chosen bunch and extraction of the optics
parameters. At the end, the machine optics is represented by a set of 6D linear
maps between the interaction points.

It was estimated that resonances generated by known Tevatron nonlinearities,
such as the final focus triplets and lattice sextupoles, were much weaker than
those due to beam-beam collisions at the operational betatron tune working
point. Hence, inclusion of nonlinear lattice elements into the simulation was
deemed unnecessary. Still, the code has the capability to include thin
multipoles up to the 10-th order.

\subsection{Chromaticity}
Although linear optics is used for the machine lattice model, there are two
nonlinear lattice effects which are considered to be significant for beam-beam
behaviour and were included into simulations. These are the chromaticities of
beta-functions excited in the main IPs and chromaticities of the betatron tunes.
In the Hamiltonian theory the chromaticity of beta-functions does not come from
energy-dependent focusing strength of quadrupole magnets
(as one would intuitively expect)
but from drift spaces where the transverse momentum is large (low-beta regions).
The symplectic transformations for that are:
\begin{eqnarray}
   X & = & X - L \cdot X^{\prime} \cdot \frac{\Delta p}{p} \nonumber \\
   Y & = & Y - L \cdot Y^{\prime} \cdot \frac{\Delta p}{p} \nonumber \\
   Z & = & Z - L \cdot (X^{\prime 2} + Y^{\prime 2}) / 2   \nonumber
\end{eqnarray}
\noindent
where $X$, $Y$, and $Z$ are the particle coordinates, and $L$ is the
``chromatic drift'' length. Then, it is necessary to adjust the
betatron tune chromaticities which are also affected by ``chromatic drift''.
For that, an artificial element (insertion) is used with the following
Hamiltonian:
$$   H = I_x \cdot (2 \pi Q_x + C_x \frac{\Delta p}{p}) +
       I_y \cdot (2 \pi Q_y + C_y \frac{\Delta p}{p}) ,$$

\noindent
where $I_x$ and $I_y$ are the action variables, $Q_x$ and $Q_y$ are the betatron
tunes, $C_x$ and $C_y$ are the [additions to the] chromaticities of the betatron
tunes.

\subsection{Diffusion and noise}
Diffusion and noise are simulated by a single random kick applied to the
macroparticles once per turn. Strength of the kick at different coordinates is
given by a symmetrical matrix representing correlations between Gaussian noises.
In the Tevatron operation, the diffusion was rather slow in terms of the computer
simulation -- the characteristic time for the emittance change was about an
hour, or $\sim 10^8$ turns. In our simulations of the beam-beam dynamics,
the noise was artificially increased by
three orders of magnitude in order to match the diffusion and the computer
capabilities.

We justify this approach below.
In contrast to electron-positron colliders there is no synchrotron radiation
damping in hadron colliders.
As the result, during the store time the effect of beam-beam
interaction on the emittance growth needs to be minimized and made small
relative to other diffusion mechanisms such as the intra-beam scattering (IBS),
scattering on the residual gas, and diffusion due to RF phase noise, etc. We will
call these the extrinsic diffusion to distinguish from the diffusion excited by
beam-beam effects. For the 2005 Tevatron parameters the extrinsic diffusion set
the luminosity lifetime to be about 10 hours at the beginning of the store. IBS
dominated both transverse and longitudinal diffusions in the case of protons
while its relative effect was significantly smaller for antiprotons because of
$\sim$5 times smaller intensity.

Table \ref{table1} summarizes lifetimes for major beam parameters obtained 
with diffusion model \cite{diffusion} for a typical 2005 Tevatron store with the 
luminosity of $0.9 \times 10^{32}$cm$^{-2}$s$^{-1}$. There were many parameters 
in Tevatron which are beyond our control and therefore each store was different. 
For good stores, the beam-beam effects made comparatively small contribution 
to the emittance growth yielding luminosity lifetime in the range of 7-8 hours 
and 10-15\% loss in the luminosity integral.
\begin{table}[hbt]
\begin{center}
\caption{Lifetimes for major beam parameters obtained with diffusion model.}
\begin{tabular}{|l|c|c|}
\hline
\textbf{Parameter} (lifetime, hour) & \textbf{Protons} & \textbf{Antiprotons} \\ 
\hline
Luminosity & 9.6 & 9.6 \\ \hline
Transverse emittance, & -17 / -18 & -52 / -46 \\
$(d\epsilon / dt)/\epsilon$ [hor./vert.] & & \\
\hline
Longitudinal emittance & -8 & -26 \\ \hline
Intensity & 26 & 155 \\ \hline
\end{tabular} 
\label{table1}
\end{center}
\end{table}

Under the Tevatron operational conditions, the emittance growth rate was small and
exact simulations of beam-beam effects would require tracking for billions of
turns. That is well beyond capabilities of present computers. Fortunately, the
extrinsic diffusion is much faster than the beam-beam diffusion.
That leads to the loss of phase correlation after about 50,000 turns. 

The external noise plays an important role in particle dynamics: it provides
particle transport in the regions of the phase space which are free from resonance
islands.

To make this transport faster we can artificially increase the noise level 
assuming that its effect scales as noise power multiplied by number of turns. 
If we choose it so that the noise alone gives 10\% emittance growth in $10^6$ 
turns (we use this level as the reference) than this number of turns of 
simulation will correspond to $\sim 5$ h of time in the Tevatron.

To verify this approach we studied the effect of the noise level on luminosity 
using the reconstructed optics.

\begin{figure}[htb]
\includegraphics*[width=\textwidth]{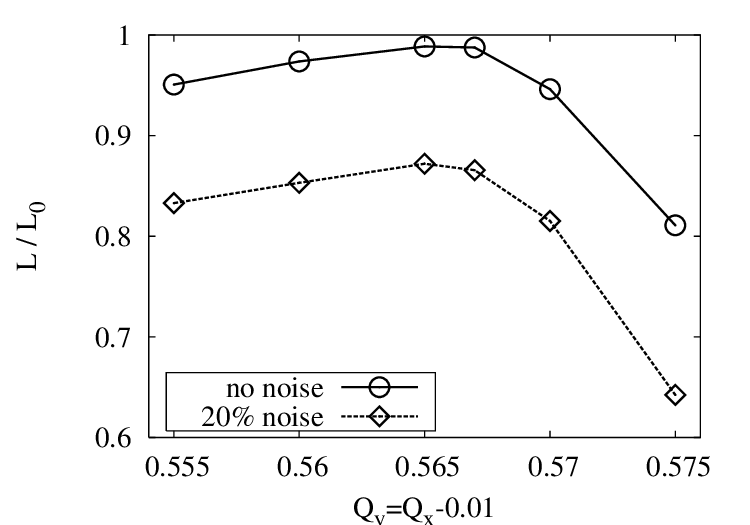}
\caption{\label{lum_tune}The reduction of luminosity after 
$t=2\cdot10^6$ turns to the initial luminosity vs. betatron tune (simulations).
Circles - no noise, diamonds - with noise amplitude corresponding to 20\%
emittance growth.}
\end{figure}

Figure \ref{lum_tune} presents the results of tune scan along the main diagonal
with the reference noise level and without noise. The effect of noise on 
luminosity corresponds to its level with exception of the point $Q_y=0.575$ 
where it was larger due to some cooperation with strong 5th order resonances.

\begin{figure}[htb]
\includegraphics*[width=\textwidth]{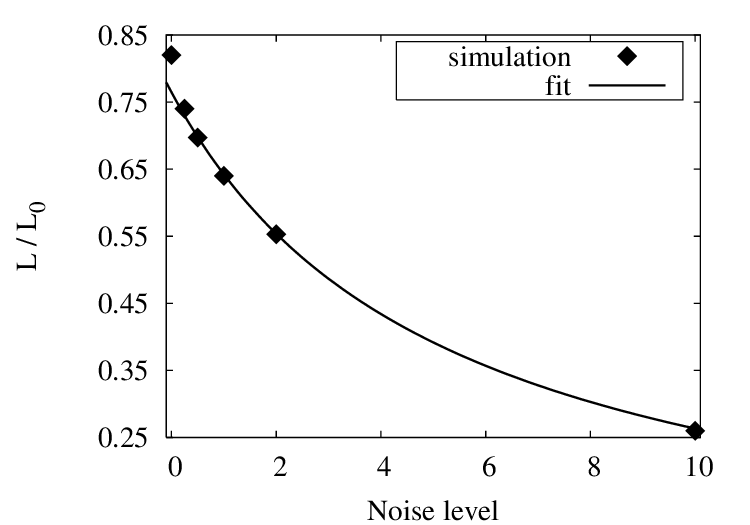}
\caption{\label{lum_noise2}Reduction of luminosity after $2\cdot 10^6$ turns vs.
noise amplitude (simulations at $Q_y=0.575$).}
\end{figure}

To study this cooperation in more detail we performed tracking at this working 
point with different noise levels. Fig.\ref{lum_noise2} shows the luminosity 
reduction in $2\cdot 10^6$ turns (diamonds) and a fit made using just 3 points, 
with relative noise level 0.5, 1 and 2.

The fit works fine for higher noise level, but predicts somewhat faster
luminosity decay in the absence of noise than actually observed in tracking.
This means that there are regions in the phase space which particles cannot pass
(within the tracking time) without assistance from the external noise so that
the simple rule for diffusion coefficiends $D$,
$D_{total}=D_{beam-beam}+D_{noise}$ does not apply. However, such
``blank spaces'' may contain isolated resonance islands which would show up on
a longer time scale with the real level of external noise.
The applicablity of this rule at the reference noise level testifies that 
(with the chosen number of turns) no such ``blank spaces'' were left so we 
get more reliable predictions.

Since 2007 we do no longer use the artificial noise enhancement for two reasons:
a) the time interval of interest became shorter (less than one hour) after the
shift of focus in beam-beam effects from antiprotons to protons; b)
over the years the available computing power was constantly increasing and
e.g. now, a full tracking over $10^7$ turns (corresponding to 210 s of real time)
takes about 20 hours on a modern computing cluster.

\subsection{Program features}
Since Lifetrac uses the ``weak-strong'' model, it can be very
efficiently parallelized. Each processor tracks its own set of particles and
the nodes need to communicate very rarely (at the end of each step), just to
gather the obtained statistics. Hence, the productivity grows almost linearly
with the number of nodes.

There are also two auxiliary GUI codes. The first one automates production
of the Lifetrac input files for different bunches from the OptiM machine lattice
files. The second one is dedicated for browsing the Lifetrac output files and
presenting the simulation results in a text and graphical (histogram) form.

\subsection{Code validation}
\begin{figure}[htb]
   \includegraphics*[width=\textwidth]{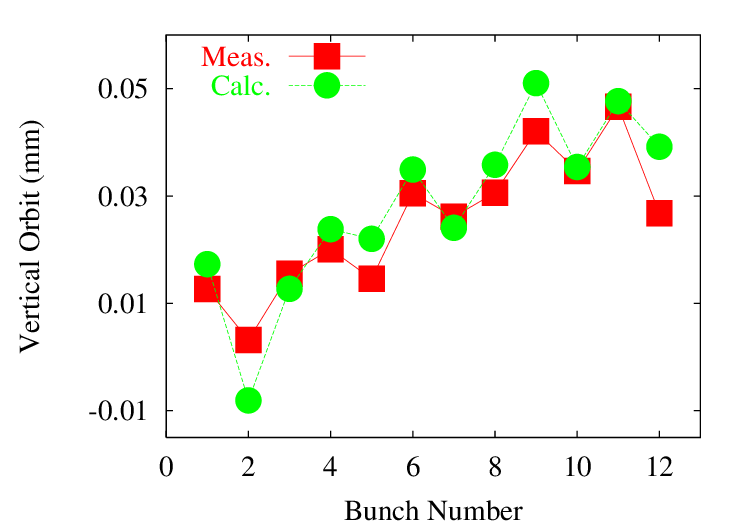}
   \caption{\label{vorb_bunch}Bunch by bunch antiproton vertical orbit:
   squares - measurements, circles - Lifetrac simulations.}
\end{figure}

\begin{figure}[htb]
   \includegraphics*[width=\textwidth]{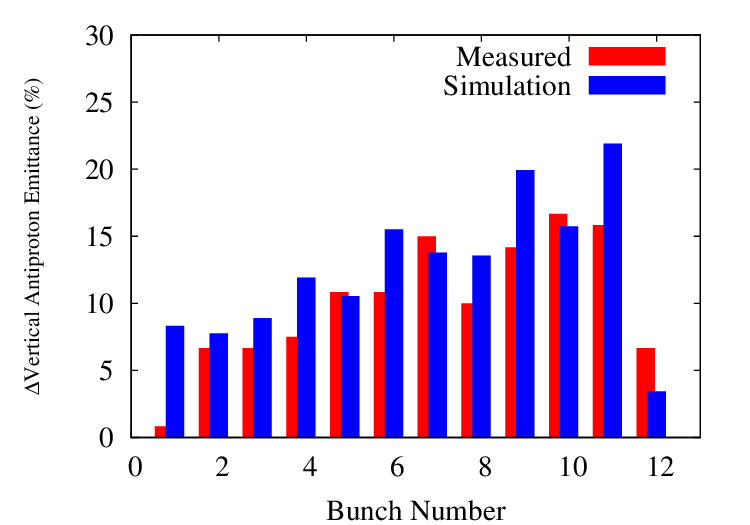}
   \caption{\label{scallops}Bunch by bunch antiproton emittance growth.
   Measured in store 3554 (red) and simulated with Lifetrac (blue).}
\end{figure}

We have validated the code using available experimental data. As an example,
Figures \ref{vorb_bunch} and \ref{scallops} show a good reproduction of the two
distinct effects in the bunch to bunch differences caused by beam-beam effects:
variation of vertical bunch centroid position due to long range dipole kicks,
and variation of transverse emittance blowup caused by difference in tunes and
chromaticities. We also demonstrated that scallops can be reduced by moving the
working point farther from 5th order resonance.

In addition, Lifetrac was validated against another weak-strong
tracking code Sixtrack on the case of the Large Hadron Collider, and good
agreement was observed \cite{SixBenchmark}.

\section{Simulation results}
Simulations with Lifetrac played an important role in justification of many
collider configuration changes, which resulted in performance improvements.
These changes include the decrease of antiproton betatron
tune chromaticity, reduction of the $\beta^*$ from 0.35 m to 0.28 m
(both in 2005), correction of the collision optics, increase of separation at
the long range collision points nearest to the main IPs, and correction of the
chromatic beta-function. In this section we present selected simulation
results for some of these topics.

\subsection{Optics errors}
\begin{figure}[htb]
\includegraphics*[width=\textwidth]{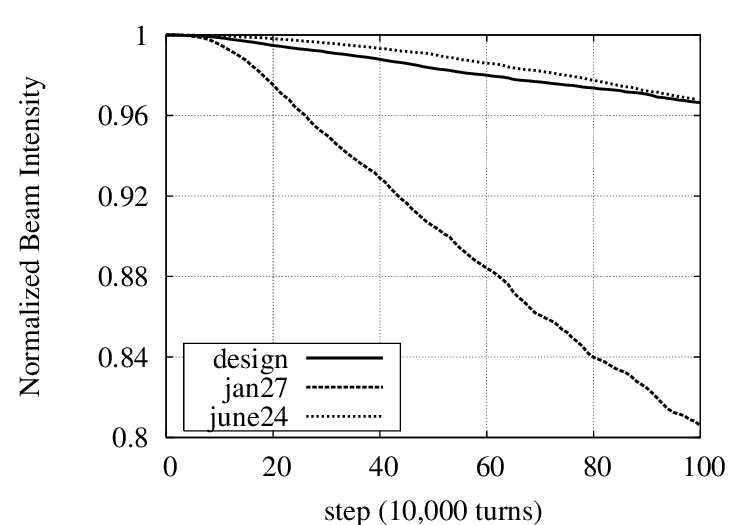}
\caption{\label{intensity1}Intensity of antiproton bunch \#6 vs. time for
different types of optics (simulations with Lifetrac).
$\xi=0.01$, $Q_x=0.57$, $Q_y=0.56$}
\end{figure}

Early in the Collider Run II it was recognized that the Tevatron collision
optics had significant deviations from design caused by the systematic betatron
coupling resulting from the coil creep in main dipoles \cite{TevDipoles},
magnet calibration errors and other sources. We have measured the machine optics
using the orbit response method \cite{diff_orb} and performed simulations with
Lifetrac for different
optics versions. In the results presented below we used 3 major optics
modifications:
\begin{itemize}
\item ``design'' optics with ideal parameters of the main IPs and zero coupling.
\item ``january'' optics which was in effect until March, 2004.
This optics was measured in January, 2004, and had sufficient distortions
in the main IPs (unequal $\beta$-functions, beam waists shifted from the IP),
and betatron coupling.
\item ``june'' optics introduced in March, 2004, where all the distortions were
corrected.
\end{itemize}
Comparison of the three cases is shown in Figure \ref{intensity1}. 
This plot shows that modifications to the optics implemented in March, 2004,
made the optics close to the design. Additional simulations revealed that the
main source of particle losses was in the long range collisions nearest
to the main IPs. Increasing the beams separation in those points and optimisation
of the phase advances cured high antiproton losses.

\subsection{\label{sec-colhelix}New collision helical separated orbit}
\begin{figure}[htb]
\includegraphics*[width=\textwidth]{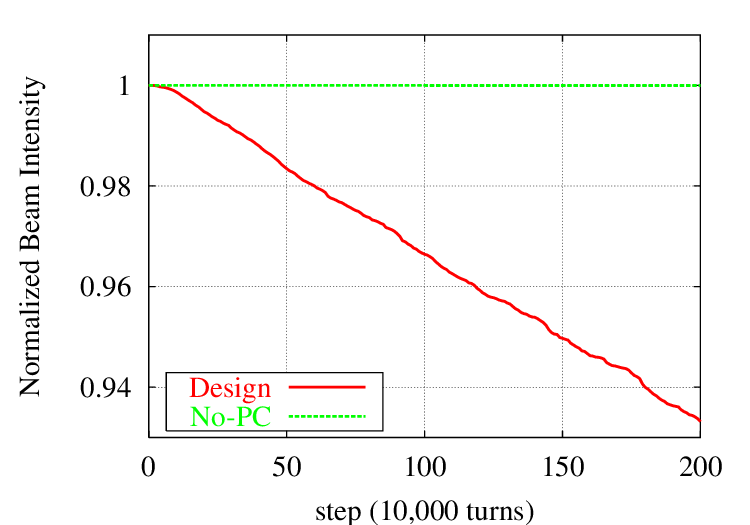}
\caption{\label{fig_nopc}Normalized intensity of antiproton bunch \#6 simulated
in the presence
(solid line) and in the absence (dashed line) of the parasitic beam-beam
interactions.}
\end{figure}

As mentioned above, the strong betatron resonances affecting the collider
performance were caused by beam-beam effects.
It was shown analytically that the strength
of the 7-th order resonance was determined by the long range collisions
\cite{YAHelices}. Our simulations predicted that increasing the beam
separation at the parasitic collision (PC) points nearest to the main IPs would
give the largest benefit.
The significance of the PCs is illustrated in Figure \ref{fig_nopc}, where
a bunch intensity is plotted vs. time ($2\times 10^6$ turns in this simulation
correspond to about 15 hours in the Tevatron) with the complete set of IPs and
PCs, and with the most significant PCs turned off. It is clear that PCs dominate
the particle losses.

To increase separation at these PCs, two extra electrostatic separators were
installed during the 2006 shutdown. As the result,
the separation at the PCs upstream and downstream of the main collision points
(CDF and D0) increased
by about 20\% (Table \ref{TableSeps}). The larger separation showed itself in
improved antiproton lifetime and allowed to push the proton intensity limit further.
\begin{table}[hbt]
\begin{center}
\caption{Radial separations in the first long range collision points in the units
of the rms beam size.}
\begin{tabular}{|l|c|c|c|c|}
\hline
 & \textbf{CDF u.s.} & \textbf{CDF d.s.} &
 \textbf{D0 u.s.} & \textbf{D0 d.s.}\\ \hline
 \textbf{Before}  & 5.4 & 5.6 & 5.0 & 5.2 \\ \hline
\textbf{After} & 6.4 & 5.8 & 6.2 & 5.6 \\ \hline
\end{tabular}
\label{TableSeps}
\end{center}
\end{table}

\begin{figure}[htb]
\includegraphics[angle=-90,width=\textwidth]{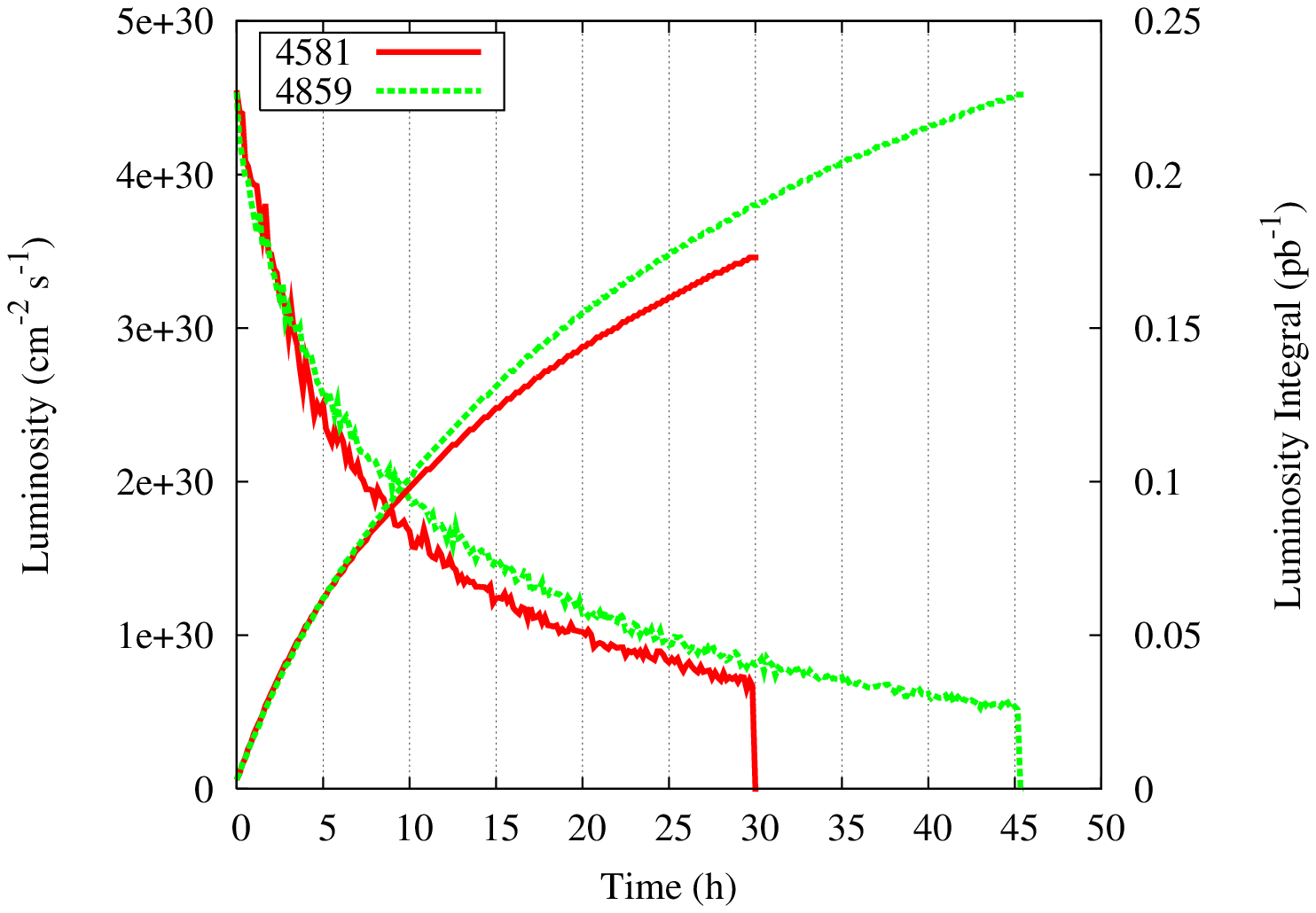}
\caption{\label{LumiComp}Single bunch luminosity and luminosity integral for
stores 4581 and 4859 (correspondingly - before and after installation of additional
electrostatic separators.}
\end{figure}

Figure \ref{LumiComp} shows a comparison of the single bunch luminosity and
luminosity integral  for two HEP stores before and after commissioning of the
new helical orbits. Initial intensities and emittances of antiprotons in these
stores were close which allows direct comparison.
As one can see, luminosity lifetime in the new configuration
has substantially improved. The overall gain can be quantified in
terms of luminosity integral over a fixed period of time (e.g. 24 hours)
normalized by the initial luminosity. The value of this parameter has
increased by 16\%.

\subsection{Betatron tune chromaticity}
\begin{figure}[htb]
\includegraphics*[width=\textwidth]{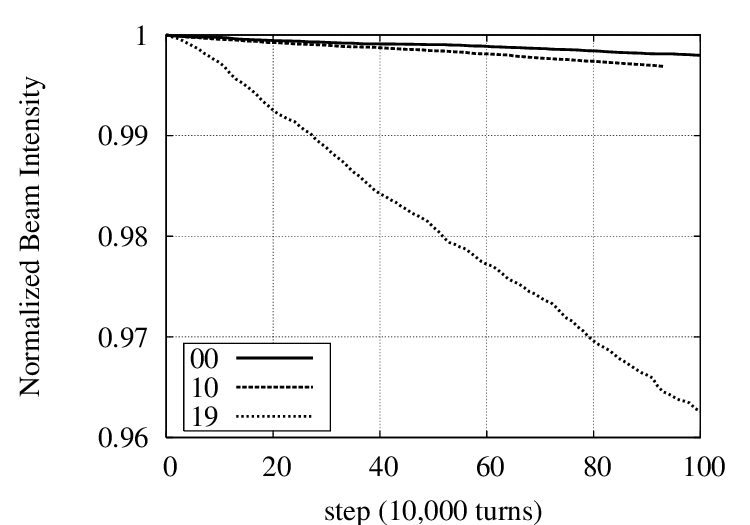}
\caption{\label{intensity2}Evolution of the antiproton bunch intensity for
various values of betatron tune chromaticity, $Q'=0,10,19$.
$Q_x=0.58$, $Q_y=0.575$, $\xi=0.01$.}
\end{figure}

Reducing the betatron tune chromaticity can also be a very powerful instrument
in decreasing the particle losses. Simulation results in Figure \ref{intensity2}
demonstrate that changing the tune chromaticity from 
15-20 units to 5-10 units may significantly improve the beam lifetime. 
This change was implemented in 2006 and resulted in about 10\% gain in the
luminosity integration rate. The safe lower limit of the tune chromaticity
was determined by the coherent stability of the beams. It was demonstrated
experimentally that with head-on collisions initiated, the beams remained
stable even at zero chromaticity. Apparently, the Landau damping by strong
nonlinearity of the head-on beam-beam was the major factor.
However, in the routine operation the typical value of chromaticity was set at
approximately 5.

\subsection{$\beta^*$ reduction}
Another improvement which could be relatively easily implemented was the 
reduction of the beta-function at the main IPs. Decreasing the $\beta^*$
from the design value of 0.35~m to 0.28~m resulted in a 10\% gain both in peak
luminosity and in the luminosity integral. However, further improvement along this
route was not practical due to the hourglass effect and rather significant
increase of the maximum beta-function in the final focus triplet, and
subsequent enhancement of effects related to the magnet vibrations and aperture
limitation.

\subsection{\label{sec-C2}Second order chromaticity}

Increasing the beam separation mitigated the long range beam-beam effects.
However, with advances in the antiproton production rate, the initial antiproton
intensity at collisions has been rising continuously. In 2006, the head-on
beam-beam parameter for protons was pushed up to 0.016 which made
the head-on beam-beam effects in the proton beam much more pronounced.
One of the possible ways for improvement was a major change of the betatron
tune in order to increase the available tune space (e.g. close to the half-integer).
That, however, would require
significant investment into the machine time for optics studies and tuning.
A partial solution could be implemented by decoupling of the transverse and
longitudinal motion at the main IPs i.e., by reducing the chromatic
beta-function.

The value of chromatic beta-function $(\Delta \beta / \beta) / (\Delta p /p)$
at both IPs in the original Tevatron lattice was 
-600 which lead to the beta-function change of 10\% for a particle with
1$\sigma$ momentum deviation \cite{C2Comp}. 
Thus, a large variation of focusing for particles in the bunch existed giving
rise to beam-beam driven synchrobetatron resonances.

\begin{figure}[htb]
   \includegraphics*[width=\textwidth]{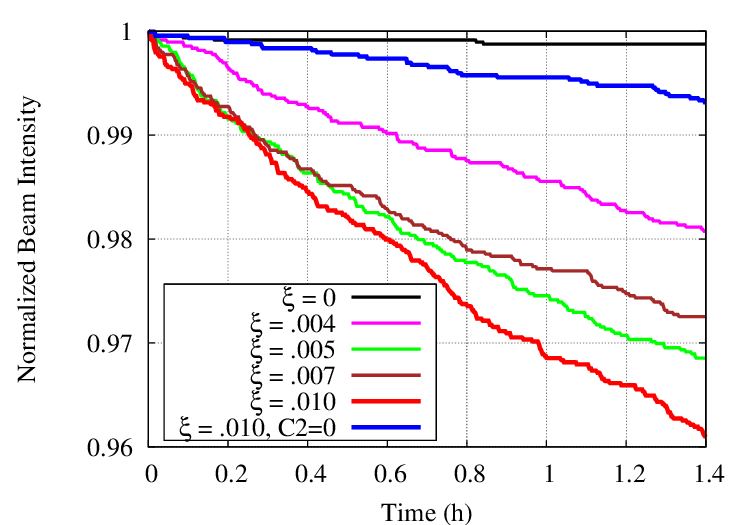}
   \caption{\label{intensity_c2}Proton intensity evolution for different values
   of the beam-beam tune shift parameter per IP from $\xi=0$ to $\xi=0.01$;
   without and with compensation of the chromaticity of $\beta^*$ ($C_2=0$).}
\end{figure}

Planning ahead for the increase in amount of antiprotons available to the collider,
we identified the large chromaticity of $\beta^*$ as a possible source of the
proton lifetime deterioration. Figure \ref{intensity_c2} shows the beam-beam
induced proton lifetime for different values of $\xi$, and demonstrates the
benefitial effect of corrected chromatic $\beta^*$.

\begin{figure}[htb]
   \includegraphics*[width=\textwidth]{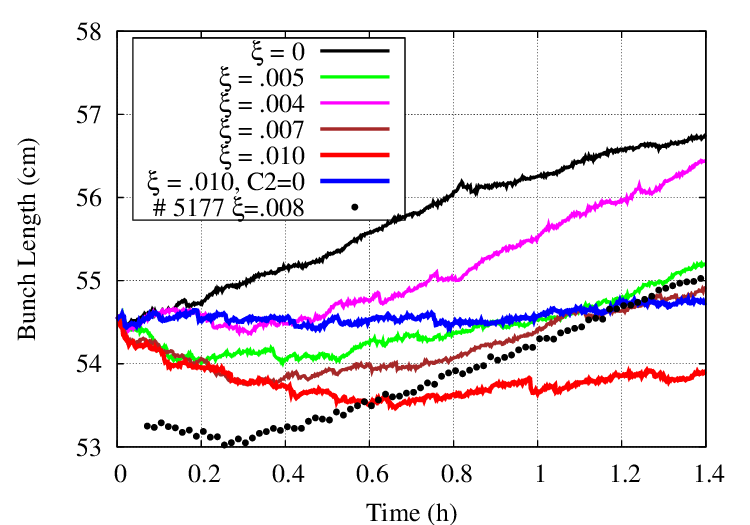}
   \caption{\label{sigm_c2}Effect of the corrected second order chromaticity
   ($C_2=0$) on
   the proton bunch length evolution at different values of the beam-beam
   parameter (solid lines - simulations, dots - Tevatron HEP store \#5177).}
\end{figure}

Simulations revealed an interesting feature in the behavior of the proton bunch
length at high values of $\xi$ -- the so-called ``bunch shaving'', when the bunch
length starts to decrease after initiating head-on collisions instead of steady
growth predicted by the diffusion model (Figure \ref{sigm_c2}). This behavior was
observed multiple times during HEP stores in 2007, being especially pronounced
when the vertical proton betatron tune was set too high.

In order to achieve the desired smaller beta-function chromaticity, a new
scheme of sextupole correctors in the Tevatron has been developed and
implemented in May of 2007. The scheme used the existing sextupole magnets,
which were
split into multiple families instead of just two original SF and SD circuits.
The effect of introducing the new circuits is illustrated in Figure \ref{lumi_int}. 

\section{Summary and discussion}
Over the last four years of the Collider Run II, Tevatron routinely operated
at the values of
head-on beam-beam tune shift for both proton and antiproton beams exceeding 0.02.
The transverse emittance of antiprotons was a factor of 3 to 5 smaller than the
proton emittance. This created significantly different conditions for the two
beams.

Beam-beam effects in antiprotons were dominated by long range interactions at
four parasitic collision points with minimal separation. After the separation at
these points was increased to 6$\sigma$ no adverse effects were observed in
antiprotons at nominal proton intensitites.

On the contrary, protons experienced intensity lifetime degradation due to head-on
collisions with the beam of smaller transverse size. Correction of chromatic
$\beta$-function in the final focus and reduction of betatron tune chromaticity
increased the dynamic aperture and improved proton beam lifetime.

Weak-strong simulation of the beam-beam effects in the Tevatron with Lifetrac code
correctly describes many observed features of the beam dynamics,
has predictive power and has been used to initiate modifications of the machine
configuration.

Further increase of the beam intensities was limited by the space available on
the tune diagram near the operational working point. A change of the tune working
point from 0.58 to near the half integer resonance would allow as much as 30\%
increase of intensities but would require a lengthy commissioning period which
rendered this improvement impossible in the Tevatron Collider Run II.

\begin{acknowledgments}
We would like to thank the Tevatron Department staff for their constant support
and help in carrying out the beam expermients. We are particularly grateful to
J.~Annala, B.~Hanna, R.S.~Moore, V.~Shiltsev, D.~Still and C.Y.~Tan.
We would like to acknowledge the assistance of computing cluster administrator
M.~Kriss. We are indebted to T.~Bolshakov for his help with the design of computer
simulations. Fermilab is operated by Fermi Research Alliance, LLC under Contract
No.~DE-AC02-07CH11359 with the United States Department of Energy.
\end{acknowledgments}


\begin{thebibliography}{99}
\bibitem{TevJINST}
S. Holmes, R.S. Moore, and V. Shiltsev,
\emph{Overview of the Tevatron collider complex: goals, operations and
performance}, \jinst{6}{2011}{T08001}.

\bibitem{eCool}
S. Nagaitsev et al., \emph{Experimental demonstration of relativistic
electron cooling}, \emph{Phys. Rev. Lett.} {\bf 96} (2006) 044801.

\bibitem{BBTev}
V. Shiltsev et al., \emph{Beam-beam effects in the Tevatron},
\emph{Phys. Rev. ST Accel. Beams} {\bf 8} (2005) 101001.

\bibitem{headtail03}
P.M. Ivanov et al., \emph{Head-tail instability at Tevatron},
in proceedings of \emph{the 2003 IEEE Particle Accelerator Conference},
Portland, OR, USA, (2003), p. 3062.

\bibitem{headtail05}
P.M. Ivanov et al., \emph{Landau damping of the weak head-tail instability at
Tevatron}, in proceedings of \emph{the 2005 IEEE Particle Accelerator
Conference}, Knoxville, TN, USA, (2005), p. 2714.

\bibitem{headtail08}
V.H. Ranjbar and P.M. Ivanov,
\emph{Chromaticity and wakefield effect on the transverse motion of longitudinal
bunch slices in the Fermilab Tevatron}, \emph{Phys. Rev. ST Accel. Beams}
{\bf 8} (2008) 084401.

\bibitem{helix05}
R.S. Moore et al., \emph{Improving the Tevatron collision helix},
in proceedings of \emph{the 2005 IEEE Particle Accelerator Conference},
Knoxville, TN, USA, (2005), p. 1931.

\bibitem{YAHelices}
Y. Alexahin, \emph{Optimization of the helical orbits in the Tevatron},
in proceedings of \emph{the 2007 IEEE Particle Accelerator Conference},
Albuquerque, NM, USA, (2007), p. 3874.

\bibitem{PBJpaper}
C.Y. Tan and J. Steimel, \emph{Controlled emittance blow up in the Tevatron},
in proceedings of \emph{the 2009 IEEE Particle Accelerator Conference},
Vancouver, BC, Canada, (2009), p. 1668.

\bibitem{C2Comp}
A. Valishev et al., \emph{Correction of second order chromaticity at Tevatron},
in proceedings of \emph{the 2007 IEEE Particle Accelerator Conference},
Albuquerque, NM, USA, (2007), p. 3922.

\bibitem{lumevol}
V. Shiltsev and E. McCrory, \emph{Characterizing luminosity evolution in the
Tevatron}, 
in proceedings of \emph{the 2005 IEEE Particle Accelerator Conference},
Knoxville, TN, USA, (2005), p. 2536.

\bibitem{syphers}
M. Syphers, \emph{Beam-beam tune distributions with differing beam sizes},
Fermilab internal report Beams Doc. 3031, (2008).

\bibitem{YAbb}
Yu. Alexahin, \emph{Theory and reality of beam-beam effects at hadron
colliders}, 
in proceedings of \emph{the 2005 IEEE Particle Accelerator Conference},
Knoxville, TN, USA, (2005), p. 544.

\bibitem{val03}
V. Lebedev, \emph{Beam physics at Tevatron complex},
in proceedings of \emph{the 2003 IEEE Particle Accelerator Conference}, 
Portland, OR, USA, (2003), p. 29.

\bibitem{storeanalysis}
A. Valishev, \emph{Tevatron store analysis package},
Software code supported in 2006-2011,
\begin{verbatim}
http://www-bd.fnal.gov/SDAViewersServlets/valishev_sa_catalog2.html
\end{verbatim}

\bibitem{lifetrac96}
D. Shatilov, \emph{Beam-beam simulations at large amplitudes and lifetime
determination}, \emph{Part. Accel.} {\bf 52} (1996) p. 65.

\bibitem{sympl}
K. Hirata, H. Moshammer and F. Ruggiero,
KEK Report 92-117, (1992).

\bibitem{diff_orb}
V. Sajaev et al., \emph{Fully coupled analysis of orbit response matrices at the
FNAL Tevatron}, in proceedings of \emph{the 2005 IEEE Particle Accelerator
Conference}, Knoxville, TN, USA, (2005), p. 3662.

\bibitem{TevOptics05}
A. Valishev et al., \emph{Progress with collision optics of the Fermilab
Tevatron collider}, in proceedings of \emph{the 2006 European Particle
Accelerator Conference}, Edinburgh, Scotland, (2006), p. 2053.

\bibitem{loco2006}
V. Lebedev et al., \emph{Measurement and correction of linear optics and
coupling at tevatron complex}, \emph{Nucl. Instrum. Methods Phys. Res., Sect. A}
{\bf 558}, (2006), p. 299.

\bibitem{optim}
V. Lebedev, \emph{OptiM code}, Private communication,
\begin{verbatim}
http://www-bdnew.fnal.gov/pbar/organizationalchart/lebedev/OptiM/optim.htm
\end{verbatim}

\bibitem{lebedev-bogacz}
V. Lebedev and A. Bogacz, \emph{Betatron motion with coupling of horizontal and
vertical degrees of freedom}, \jinst{5}{2010}{P10010}.

\bibitem{diffusion}
V. Lebedev and A. Burov, \emph{Collective instabilities in the Tevatron complex},
in proceedings of \emph{the 33rd ICFA advanced beam dynamics workshop on high
intensity and high brightness hadron beams}, Bensheim, Germany, (2004),
p. 350.

\bibitem{SixBenchmark}
F. Schmidt, A. Valishev and Y. Luo, \emph{Development and benchmarking of
codes for simulation of beam-beam effects at the LHC},
BNL Note C-A/AP/443, (2011).

\bibitem{TevDipoles}
G.E. Annala, D.J. Harding and M.J Syphers, \emph{Coil creep and skew-quadrupole
field components in the Tevatron}, \jinst{7}{2012}{T03001}.

\bibitem{IBS_JINST}
V. Shiltsev and A. Tollestrup, \emph{Emittance growth mechanisms in the
Tevatron beams}, \jinst{6}{2011}{P08001}.

\end{thebibliography}
\end{document}